\documentclass{jfm}

\usepackage{graphicx}
\usepackage{natbib}

\ifCUPmtlplainloaded \else
  \checkfont{eurm10}
  \iffontfound
    \IfFileExists{upmath.sty}
      {\typeout{^^JFound AMS Euler Roman fonts on the system,
                   using the 'upmath' package.^^J}%
       \usepackage{upmath}}
      {\typeout{^^JFound AMS Euler Roman fonts on the system, but you
                   dont seem to have the}%
       \typeout{'upmath' package installed. JFM.cls can take advantage
                 of these fonts,^^Jif you use 'upmath' package.^^J}%
      }
  \else
  \fi
\fi

\ifCUPmtlplainloaded \else
  \checkfont{msam10}
  \iffontfound
    \IfFileExists{amssymb.sty}
      {\typeout{^^JFound AMS Symbol fonts on the system, using the
                'amssymb' package.^^J}%
       \usepackage{amssymb}%
       \let\le=\leqslant  
         
      }{}
  \fi
\fi

\ifCUPmtlplainloaded \else
  \IfFileExists{amsbsy.sty}
    {\typeout{^^JFound the 'amsbsy' package on the system, using it.^^J}%
     \usepackage{amsbsy}}
    {\providecommand\boldsymbol[1]{\mbox{\boldmath $##1$}}}
\fi

\newsavebox{\astrutbox}
\sbox{\astrutbox}{\rule[-5pt]{0pt}{20pt}}

\usepackage{graphicx}% Include figure files
\usepackage{dcolumn}% Align table columns on decimal point
\usepackage{bm}% bold math
\usepackage{amssymb,amsbsy,times,fancyhdr,color}
\usepackage{amsmath}

\newcommand{\bfzhat}{\mbox{\boldmath $ {\hat z} $ } }
\newcommand{\bfxhat}{\mbox{\boldmath $ {\hat x} $ } }

\newcommand{\bfB}{\mbox{\boldmath$B$}}

\newcommand\calE{{\mathcal E}}
\newcommand\calP{{\mathcal P}}
\newcommand\bfcalE{\boldsymbol{\calE}}

\newcommand{\bfalpha}{\boldsymbol{\alpha}}

\newcommand{\bfb}{\mbox{\boldmath$b$}}

\newcommand{\bfe}{\mbox{\boldmath$e$}}
\newcommand{\bfu}{\mbox{\boldmath$u$}}
\newcommand{\bfU}{\mbox{\boldmath$U$}}

\newcommand\Ray{\mbox{\textit{Ra}}}  % Rayleigh number
\newcommand\Tay{\mbox{\textit{Ta}}}  % Taylor  number
\newcommand\kcut{k_{\rm cut}}

\title[Dynamos in Rotating Sheared Convection]{The Effect of Velocity Shear on Dynamo Action Due to Rotating Convection}

\author[D.W. Hughes and M.R.E. Proctor]%
{D.\ns W.\ns H\ls U\ls G\ls H\ls E\ls S$^1$
\and M.\ns R.\ns E.\ns P\ls R\ls O\ls C\ls T\ls O\ls R$^2$}

\affiliation{$^1$Department of Applied Mathematics, University of Leeds, Leeds LS2 9JT, UK\\[\affilskip]
$^2$Centre for Mathematical Sciences, University of Cambridge, Wilberforce Road, Cambridge CB3 0WA, UK}

\pubyear{2012}
\volume{650}
\pagerange{119--126}
\date{?; revised ?; accepted ?. - To be entered by editorial office}
\date{\today}% It is always \today, today,
\begin{document}

\maketitle

\begin{abstract}
Recent numerical simulations of dynamo action resulting from rotating convection have revealed some serious problems in applying the standard picture of mean field electrodynamics at high values of the magnetic Reynolds number, and have thereby underlined the difficulties in large-scale magnetic field generation in this regime. Here we consider kinematic dynamo processes in a rotating convective layer of Boussinesq fluid with the additional influence of a large-scale horizontal velocity shear. Incorporating the shear flow enhances the dynamo growth rate and also leads to the generation of significant magnetic fields on large scales. By the technique of spectral filtering, we analyse the modes in the velocity that are principally responsible for dynamo action, and show that the magnetic field resulting from the full flow relies crucially on a range of scales in the velocity field. Filtering the flow to provide a true separation of scales between the shear and the convective flow also leads to dynamo action; however, the magnetic field in this case has a very different structure from that generated by the full velocity field. We also show that the nature of the dynamo action is broadly similar irrespective of whether the flow in the absence of shear can support dynamo action.
\end{abstract}


\section{Introduction}
\label{sec:intr}

One of the outstanding theoretical problems in astrophysical MHD is to account for the generation of global scale magnetic fields, as detected  in many cosmic bodies. These are generally held to be produced by a hydromagnetic dynamo process, in which the magnetic fields are maintained against resistive effects by induction due to the plasma motions. It is, however, hard to provide  a convincing theoretical explanation of how such \textit{large-scale} fields--- i.e.\ those with a significant component on scales much larger than that of the plasma motions responsible for their generation ---  can be generated.

The traditional theoretical approach to explaining the generation of large-scale magnetic fields is 
via mean field electrodynamics \citep[see, for example,][]{Moffatt_78,KR_80}. Here, the evolution of the mean (large-scale) field is governed by the mean induction equation,
\begin{equation}
\frac{\partial \bfB_0}{\partial t} = \nabla \times \left( \bfU_0 \times \bfB_0 \right) + 
\nabla \times \bfcalE + \eta \nabla^2 \bfB_0 ,
\label{eq:mean_ind}
\end{equation}
where $\bfB_0$ represents the mean magnetic field, $\bfU_0$ the mean velocity, $\bfcalE$ the
mean electromotive force (emf) and $\eta$ the magnetic diffusivity. The mean emf is defined by
\begin{equation}
\bfcalE = \langle \bfu \times \bfb \rangle ,
\label{eq:mean_emf}
\end{equation}
where $\bfu$ and $\bfb$ represent the (small-scale) fluctuating velocity and magnetic fields, and angle brackets denote a spatial average over intermediate scales.  The theory proceeds on the assumption that $\bfcalE$ is a linear functional of $\bfB_0$, which leads to the expansion
\begin{equation}
\calE_i = \alpha_{ij}B_{0j} + \beta_{ijk} \frac{\partial B_{0j}}{\partial x_k} + \cdots \ .
\label{eq:emf_expansion}
\end{equation}
(\citet{HP_10} discuss the implications of a more general expansion procedure involving also temporal derivatives of the mean field.) In the kinematic regime, in which the field is assumed to exert no back-reaction on the flow, the components $\alpha_{ij}$ and $\beta_{ijk}$ depend solely on the properties of the velocity field and on the magnetic diffusivity. The symmetric part of the $\bfalpha$ tensor (the so-called `$\alpha$-effect') leads to field amplification, and can be non-zero only in flows that lack reflectional symmetry, such as helical flows. For isotropic turbulence,  $\beta_{ijk} = \beta \epsilon_{ijk}$ and the scalar $\beta$ can then be identified as a turbulent diffusivity; in general though, $\beta_{ijk}$ has a much more complicated interpretation \citep[see][]{KR_80}. In most astrophysical applications, the mean field is considered to be axisymmetric; it can then be written, in cylindrical polar coordinates $(s, \phi, z)$, as $\bfB_0 = \nabla \times \left( A(s, z)\bfe_\phi \right) + B(s, z) \bfe_\phi$.  Under the strongest simplifying assumptions of isotropic turbulence and an azimuthal mean flow, the mean field is then described by the following two equations:
\begin{align}
\frac{\partial A}{\partial t}  &= \alpha B + {\tilde \eta}(s, z) \left( \nabla^2 - \frac{1}{s^2} \right) A , \\
\frac{\partial B}{\partial t}  &= s \left( \bfB_P \cdot \nabla \right) \omega + 
\left( \nabla \times ( \alpha \bfB_P ) \right) \cdot \bfe_\phi + 
{\tilde \eta} \left( \nabla^2 - \frac{1}{s^2} \right) B +
\frac{1}{s} \nabla {\tilde \eta} \cdot \nabla(sB),
\label{eq:mean_ind_scalars}
\end{align}
where $\bfB_P$ denotes the poloidal field, $s \omega(s,z) \bfe_\phi$ is the mean flow and ${\tilde \eta}(s, z) = \eta + \beta$ is the total magnetic diffusivity. It is necessary for dynamo action that the coupling terms in these equations are non-zero; the dynamo cycle depends crucially on the generation of poloidal field from toroidal, and, conversely, the generation of toroidal field from poloidal. The former requires the $\alpha$-effect, whilst the latter can result from either the $\alpha$-effect or from shearing of poloidal field by the differential rotation, the $\omega$-effect. The resulting dynamos are designated, respectively, as $\alpha^2$ or $\alpha \omega$-dynamos. Closed form expressions for $\alpha$ and $\beta$ can be obtained only under simplifying assumptions, notably small values of the magnetic Reynolds number on the fluctuating scale, or short correlation times for the flow. Neither of these applies, however, in the astrophysical context. In consequence, astrophysical modelling typically involves adopting plausible, albeit arbitrary, spatial forms and amplitudes of $\alpha$ and $\beta$.

Recent research has focused on attempts to measure the $\alpha$-effect directly in numerical simulations of turbulent flows, either in forced helical turbulence \citep[e.g.][]{CH_96, Brandenburg_01, CHT_02} or in rotating turbulent convection \citep[e.g.][]{CH_06, HC_08}.  Mean field coefficients can only be properly determined if there is adequate separation between the small scales of the turbulence and the system size; for convective turbulence, this is most readily accomplished by adopting the relatively simple system of plane layer, Boussinesq convection. The study of large-scale dynamo action in this system has quite a long history, dating back to the pioneering papers of \cite{CS_72} and \cite{Soward_74}. Subsequently, there have been a number of numerical investigations of the problem \citep[e.g.][]{StPierre_93, JR_2000, RJ_02, SH_04}. These have considered magnetic field generation in domains with an $O(1)$ aspect ratio, driven by mildly supercritical convection at fairly rapid rotation rates; the resulting dynamo can then be interpreted as a mean field $\alpha^2$ dynamo.

More recently, with ever-improving computational performance, it has become possible to investigate  more turbulent regimes at larger aspect ratios \citep[e.g.][]{CH_06, KKB_10}. The paper of \cite{CH_06} suggested two significant problems with the standard mean field picture. The first is that when the convection is sufficiently vigorous, it acts as a \textit{small-scale} dynamo, despite the flow being significantly helical. There is no tendency to generate large-scale field; indeed, the spectrum of magnetic energy is essentially identical to that resulting from the small-scale dynamo generated by turbulent \textit{non-rotating} convection \citep{Cattaneo_99, CH_06}. Attempts to measure the $\alpha$-effect by imposing a uniform horizontal field for turbulent convection just below the dynamo threshold, but still at high magnetic Reynolds number $Rm$, reveal the second problem. Despite averages being taken over many convective cells, the $\alpha$-effect exhibits significant temporal variations about a mean value that is much smaller than the characteristic speeds of the flow. Surprisingly, a coherent helicity distribution does not lead to a significant $\alpha$-effect.

However, rather different conclusions were reported for the convective dynamo simulations of \cite{KKB_10}, who obtained significant mean fields and sizeable $\alpha$-effects. The differences can be attributed to a number of factors: ($a$) the $\alpha$-effect depends on the horizontal correlations of the turbulence --- which, in turn, depend on the degree of supercriticality and the rotation rate; ($b$) different calculations employ different magnetic boundary conditions; ($c$) there are different definitions of the $\alpha$-effect and the means of measuring it --- e.g.\ the traditional imposed field method \citep{Moffatt_78}, the test field method \citep{Schrinner_et_al_2007} and the `resetting' method \citep{Ossendrijver_et_al_2002}. A detailed discussion of all of these issues can be found in \cite{HPC_11}.

Although the precise nature of the dynamo mechanism in these simulations remains uncertain, there is no doubt that large-scale fields are observed in nature and it is therefore important to identify other mechanisms that may lead to such fields. Since most astrophysical bodies possess a large-scale differential rotation, it is natural to incorporate a large-scale shear flow into the convection model and explore the consequences for any dynamo action. In this paper we carry out such a programme, building on the results of \cite{HP_09} and \cite{PH_11}, who were the first to show that a combination of small-scale convection and large-scale velocity shear could lead to magnetic field growth on large scales. In order to obtain scale separation between the convection and the shear flow, we consider a uni-directional horizontal flow, dependent only on the other horizontal direction. We concentrate in this paper solely on the kinematic dynamo problem, in which the back-reaction of the magnetic field on the velocity via the Lorentz force is ignored; thus we examine in some detail the nature of the generation mechanism, but we do not address the means by which magnetic field growth is saturated.

The paper is organized as follows. In \S\,\ref{sec:influence_shear} we discuss the various ways in which a large-scale shear flow may affect the dynamo process. The mathematical formulation of the problem is contained in \S\,\ref{sec:formulation}. In \S\,\ref{sec:ffrsc} we consider the introduction of a velocity shear into a convective flow that, in the absence of shear, does not act as a dynamo; we describe first the characteristics of the flow and then those of the magnetic field that it generates. Section~\ref{sec:ff} looks in more detail at the dynamo process, through considering `filtered' flows, in which only certain scales in the velocity are retained. In \S\,\ref{sec:hrn} we consider a more vigorous convective state than in \S\,\ref{sec:ffrsc}, one that supports dynamo action even in the absence of shear, in order to determine whether this is a crucial factor in the nature of the ensuing dynamo action. In the concluding \S\,\ref{sec:disc} we discuss the implications of our results and their relation to parallel studies of dynamos driven by a combination of forced turbulence and uniform shear.

\section{The Influence of Velocity Shear on Convective Dynamos} \label{sec:influence_shear}

Before describing our results it is instructive to consider the various possible ways in which a large-scale velocity shear may influence the nature of dynamo action driven by rotating convection. A number of possibilities suggest themselves.

At high $Rm$, rotating convective turbulence, in the absence of shear, can induce large local emfs, but these are decorrelated in space and time, leading to a small net $\alpha$-effect. In consequence, any dynamo field generated in extended domains is predominantly small-scale \citep{CH_06}. It is though conceivable that a coherent large-scale shear may impose more order on the correlations and, in so doing, enhance the $\alpha$-effect. Alternatively, even if the mean emfs remain very small, a large shear may be able to compensate for a feeble $\alpha$-effect (or a more complicated mean-field process) to make a viable two-scale dynamo; in the classical mean field picture it is the product of $\alpha$ and $\omega$ that controls the efficiency of the dynamo.

A rather different possibility is that enhanced dynamo action may depend on the interaction of a wide range of scales, from the largest scale of the shear to the convective cell size. In an extreme version of this, dynamo action might result solely from the interactions between the large-scale shear and the  induced motions on a similarly large scale; this would then be effectively a small-scale (i.e.\ one-scale) dynamo, but on the scale of the shear flow rather than that of the convection. We shall interpret our findings with these possibilities in mind, considering cases for which the convection in the absence of shear does, and does not, act as a dynamo.

\section{Formulation}

\label{sec:formulation}

Following \cite{CH_06} and \cite{HC_08}, we consider thermally driven convection in a three-dimensional, Cartesian layer ($0 < x, y < \lambda d$, $0< z <d$)  of Boussinesq fluid rotating about the vertical. The layer has angular velocity $\Omega$, density $\rho$, kinematic viscosity $\nu$, thermal diffusivity $\kappa$ and magnetic diffusivity $\eta$. 
This basic model is then extended by the inclusion of a horizontal flow of the (dimensional) form
\begin{equation}
\bfU_0 = U_0 f(y/d) \bfxhat , \mbox{  where  }f(y/d)=\cos \frac{2 \pi y}{\lambda d} ,
\label{eq:shear}
\end{equation}
where the total velocity is now $\bfu + \bfU_0$; this is
accomplished by replacing $\bfu$ with $\bfu + \bfU_0$ in the governing equations except for the viscous term (equivalent to forcing the flow via the momentum equation, but eliminating viscous transients). We adopt a periodic flow for consistency with the periodic horizontal boundary conditions adopted in \cite{CH_06}. For the purposes of this paper we shall restrict attention to kinematic dynamo action, so that the back-reaction of the Lorentz forces on the convection is neglected, as is appropriate for very weak fields.

Following standard practice, we adopt the layer depth $d$, the thermal relaxation time $d^2/\kappa$, and the temperature drop across the layer $\Delta T$ as the units of length, time, and temperature respectively. All velocities are scaled with $\kappa /d$; in particular, $U_0$ below is now dimensionless. The governing non-dimensional equations for the velocity $\bfu$, temperature perturbation $\theta$ and magnetic field $\bfB$ can then be expressed as
\begin{equation}
(\partial_t - \sigma \nabla^2) \bfu + \bfu \cdot \nabla \bfu +
U_0 \left( f(y) \partial_x \bfu + f'(y) u_y \bfxhat \right) +\sigma \Tay^{1/2}  \bfzhat \times \bfu =
-\nabla p  +\sigma \Ray\,\theta\bfzhat ,
\label{eq:mom}
\end{equation}
\begin{equation}
(\partial_t - \zeta \nabla^2) \bfB + \bfu \cdot \nabla \bfB +
U_0 f(y) \partial_x \bfB = \bfB \cdot \nabla \bfu + U_0 f'(y)B_y \bfxhat ,
\label{eq:ind}
\end{equation}
\begin{equation}
(\partial_t - \nabla^2) \theta + \bfu \cdot \nabla \theta +
U_0 f(y)\partial_x \theta = \bfu \cdot \bfzhat ,
\label{eq:energy}
\end{equation}
\begin{equation}
\nabla \cdot \bfB = \nabla \cdot \bfu = 0 ,
\label{eq:div}
\end{equation}
where $w$ is the vertical velocity, and $\theta$ denotes the temperature fluctuations relative to a linear background profile \citep[e.g.][]{Chandra_61}. As noted in the introduction, we consider here only the kinematic dynamo problem, and thus the Lorentz force is omitted in the momentum equation~(\ref{eq:mom}); the problem is then linear in the magnetic field, the scaling of which is arbitrary. Five dimensionless parameters appear explicitly in the governing equations: the Rayleigh number $\Ray = g{\tilde \alpha} {\tilde \beta} d^4/\kappa \nu$ (where $g$ is the gravitational acceleration, $\tilde \alpha$ is the coefficient of thermal expansion and ${\tilde \beta}$ is the superadiabatic temperature gradient), which measures the strength of thermal buoyancy relative to dissipation; the Taylor number $\Tay = 4\Omega^2d^4/\nu^2$; 
the kinetic and magnetic Prandtl numbers
\begin{equation}
\sigma = \frac{\nu}{\kappa}\quad {\rm and} \quad \sigma_m = \frac{\nu}{\eta}\;,
\label{eq:prandtl}
\end{equation}
and the dimensionless speed $U_0$. Additionally there is the choice of the aspect ratio $\lambda$.

The purely hydrodynamic solution is evolved until it reaches a stationary state, starting from an initial condition of a small perturbation to the shear flow~(\ref{eq:shear}). We then consider the dynamo action resulting from such stationary states.  It should be noted that although a flow with a large-scale component (i.e.\ with the same spatial dependence as the `target flow' (\ref{eq:shear})) does indeed occur, its amplitude may differ appreciably from $U_0$; the hydrodynamic state that ensues depends on interactions between the shear flow and convection and, possibly, on instabilities of the shear flow itself. We also introduce the derived quantity
\begin{equation}
S=2\pi U_0\ell/\lambda u_{\rm rms},
\label{eq:Sdef}
\end{equation}
where $\ell$ and $u_{\rm rms}$ are estimates, respectively, of the horizontal scale of the convection and of the typical velocity in the absence of shear; $S$ provides a measure of the competition between shear and convection.

In the horizontal directions we assume that all fields are periodic with periodicity $\lambda$. In the vertical we consider standard illustrative boundary conditions on the temperature and velocity fields, namely that the boundaries are perfect thermal conductors, impermeable and stress-free. Formally these correspond to
\begin{equation}
\theta=w=\partial_z u=\partial_z v=0 \ \ \textrm{at}\ \  z=0, 1.
\label{eqn_bc1}
\end{equation}
The natural average in this system is one over horizontal planes, which involves averaging over many convective cells. From the point of view of generating large-scale fields with the simplest vertical structure, it is therefore preferable to choose perfectly conducting boundary conditions, for which the field is purely horizontal, thereby admitting field 
configurations with only one node in the vertical. Thus we choose
\begin{equation}
B_z=\partial_zB_x=\partial_zB_y=0 \ \ \textrm{at} \ \  z=0, 1.
\label{eqn_bc2}
\end{equation}

Equations (\ref{eq:mom}) -- (\ref{eq:div}) are solved numerically by standard pseudo-spectral methods optimized for machines with parallel architecture. Details concerning the numerical methods can be found in \cite{CEW_03}.

\cite{CH_06} and \cite{HC_08} explored dynamo action and mean emf generation in systems with the fixed values of $\Tay = 500\,000$, $\sigma=1$, $\sigma_m=5$ and for values of $\Ray$ between $80\,000$ (slightly above the onset of convection) and $1\,000\,000$; the onset of dynamo action is at $\Ray \approx 170\,000$. \cite{HP_09} and \cite{PH_11} examined the influence of a range of values of shear amplitude $U_0$ for the same fixed values of $\Tay$, $\sigma$ and $\sigma_m$, for $\Ray=150\,000$ (for which there is no dynamo in the absence of shear) and aspect ratio $\lambda=5$. Here we consider extended spatial domains, with $\lambda=10$ and, for a few runs, $\lambda=20$, and consider both $\Ray=150\,000$ and $\Ray=250\,000$ (for which there is a small-scale dynamo in the absence of shear). For $\Ray=150\,000$, $u_{\rm rms} \approx 60$ (in the absence of shear), and the width of the convective cells is comparable with, though a little smaller, than the layer depth; taking $l \approx d$ leads to $S\approx U_0/600$ in this case. In \S\ref{sec:ffrsc} and \S\ref{sec:ff} we concentrate on the case of $\Ray=150\,000$; similarities and differences for the case of $\Ray=250\,000$ are discussed in \S\ref{sec:hrn}. The numerical resolution and parameter values for all the simulations presented in this paper are summarized in Table~\ref{tab:sims}.

\begin{table}
  \begin{center}
\def~{\hphantom{0}}
\begin{tabular}{ccccccc}
%\hline
$Ra$    &   $Ta$   &  $U_0$ & $\sigma$ & $\sigma_m$ & $ \lambda$  & $ N_x \times N_y \times N_z$ \\
%\hline
$1.5\times 10^5$  &   $5 \times 10^5$ & $0$ & $1$ & $5$ & $10$     &   $512\times  512 \times 97$ \\
$1.5\times 10^5$  &   $5 \times 10^5$ & $100$ & $1$ & $5$ & $10$     &   $512\times  512 \times 97$ \\
$1.5\times 10^5$  &   $5 \times 10^5$ & $200$ & $1$ & $5$ & $10$     &   $512\times  512 \times 97$ \\
$1.5\times 10^5$  &   $5 \times 10^5$ & $300$ & $1$ & $5$ & $10$     &   $512\times  512 \times 97$ \\
$1.5\times 10^5$  &   $5 \times 10^5$ & $400$ & $1$ & $5$ & $10$     &   $512\times  512 \times 97$ \\
$1.5\times 10^5$  &   $5 \times 10^5$ & $500$ & $1$ & $5$ & $10$     &   $512\times  512 \times 97$ \\
$1.5\times 10^5$  &   $5 \times 10^5$ & $600$ & $1$ & $5$ & $10$     &   $512\times  512 \times 97$ \\
$1.5\times 10^5$  &   $5 \times 10^5$ & $700$ & $1$ & $5$ & $10$     &   $512\times  512 \times 97$ \\
$1.5\times 10^5$  &   $5 \times 10^5$ & $800$ & $1$ & $5$ & $10$     &   $512\times  512 \times 97$ \\
$1.5\times 10^5$  &   $5 \times 10^5$ & $900$ & $1$ & $5$ & $10$     &   $512\times  512 \times 97$ \\
$1.5\times 10^5$  &   $5 \times 10^5$ & $1000$ & $1$ & $5$ & $10$   &   $512\times  512 \times 97$ \\
$1.5\times 10^5$  &   $5 \times 10^5$ & $2000$ & $1$ & $5$ & $10$     &   $512\times  512 \times 97$ \\
$1.5\times 10^5$  &   $5 \times 10^5$ & $1000$ & $1$ & $5$ & $20$     &   $1024\times 1024 \times 97$ \\
$1.5\times 10^5$  &   $5 \times 10^5$ & $2000$ & $1$ & $5$ & $20$     &   $1024\times 1024 \times 97$ \\
$2.5\times 10^5$  &   $5 \times 10^5$ & $0$ & $1$ & $5$ & $10$     &   $512\times  512 \times 97$ \\
$2.5\times 10^5$  &   $5 \times 10^5$ & $200$ & $1$ & $5$ & $10$     &   $512\times  512 \times 97$ \\
$2.5\times 10^5$  &   $5 \times 10^5$ & $400$ & $1$ & $5$ & $10$     &   $512\times  512 \times 97$ \\
$2.5\times 10^5$  &   $5 \times 10^5$ & $600$ & $1$ & $5$ & $10$     &   $512\times  512 \times 97$ \\
$2.5\times 10^5$  &   $5 \times 10^5$ & $800$ & $1$ & $5$ & $10$     &   $512\times  512 \times 97$ \\
$2.5\times 10^5$  &   $5 \times 10^5$ & $1000$ & $1$ & $5$ & $10$     &   $512\times  512 \times 97$ \\
$2.5\times 10^5$  &   $5 \times 10^5$ & $1200$ & $1$ & $5$ & $10$     &   $512\times  512 \times 97$ \\
$2.5\times 10^5$  &   $5 \times 10^5$ & $1400$ & $1$ & $5$ & $10$     &   $512\times  512 \times 97$ \\
$2.5\times 10^5$  &   $5 \times 10^5$ & $1630$ & $1$ & $5$ & $10$     &   $512\times  512 \times 97$ \\
%\hline
\end{tabular}
  \caption{Summary of the parameter values and numerical resolution for the simulations.}
  \label{tab:sims}
  \end{center}
\end{table}

\section{Flows and Fields in Rotating Sheared Convection} \label{sec:ffrsc}

\subsection{Influence of Shear on Convection}\label{subsec:isc}

Understanding the interactions between a shear flow and rotating convection is a complex problem, of relevance for both stellar and planetary physics. The majority of work has focused on the case of a shear flow dependent on the vertical direction, arising from relative motion of the horizontal boundaries or, alternatively, from a fictitious force \citep[see, for example,][]{HTG_80,HS_83,HS_86,KP_91,MC_97,Cox_98}. The hydrodynamical problem of a horizontally dependent shear flow, the case we consider here, may also be of relevance in planetary atmospheres, and has been examined in the nonlinear regime by \cite{HS_87}.

%------------
\begin{figure}
\vskip -0.2 cm
\centerline{
		\includegraphics[scale = 0.45]{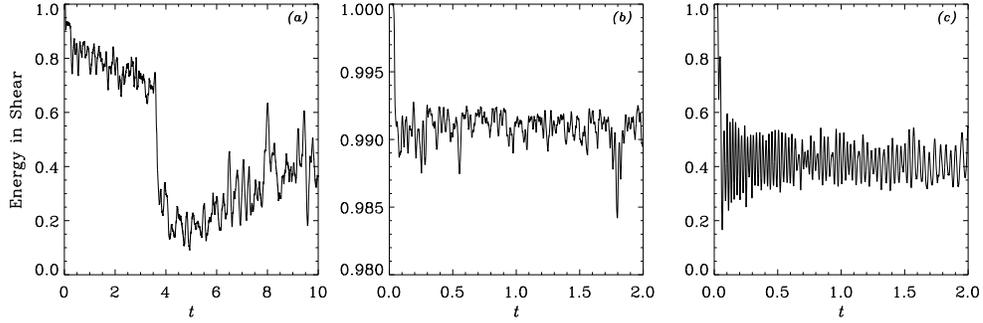}
	}
\caption{Temporal evolution of the energy in the target shear flow mode (i.e.\ the mode with wavenumbers $k_y=1$, $k_x=k_z=0$) normalized by the total kinetic energy of the flow: ($a$)~$U_0 = 300$, ($b$)~$U_0 = 1000$, ($c$)~$U_0 = 2000$.}
\label{fig:ke}
\end{figure}
%------------
\begin{figure}
\vskip -2cm
\centerline{
		\includegraphics[scale = 0.7]{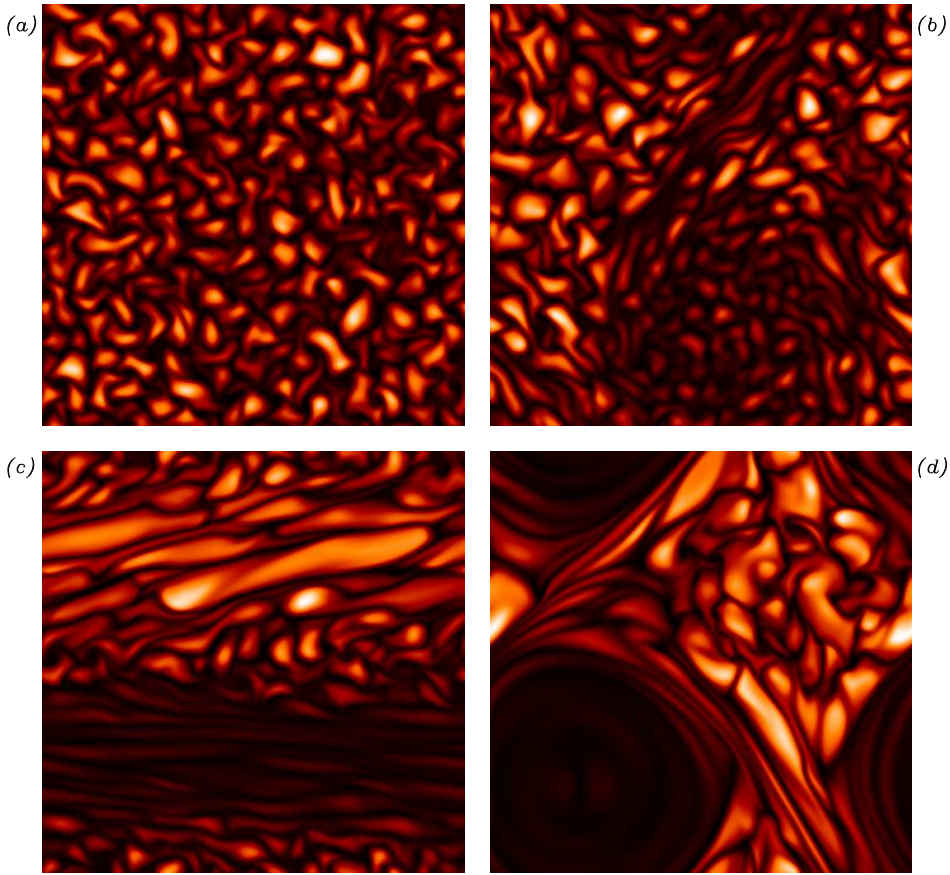}
	}
\caption{Snapshots of the temperature perturbations close to the upper boundary for $\Ray = 150\,000$ and four different values of the shear flow: ($a$) $U_0 = 0$, ($b$) $U_0 = 300$, ($c$) $U_0 = 1000$, ($d$) $U_0 = 2000$. White denotes hot fluid, black cool fluid. The target shear flow is $\bfU_0 = U_0 \cos (2 \pi y /\lambda) \bfxhat$, where $x$ is the direction of the abscissa.}
\label{fig:temps}
\end{figure}
\begin{figure}
\vskip -2.5cm
\centerline{
		\includegraphics[scale = 0.6]{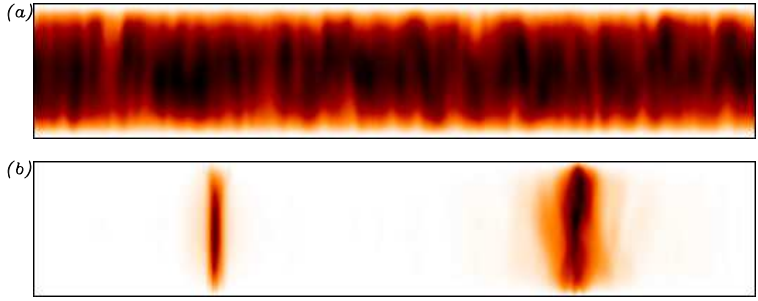}
	}
\caption{Snapshots of the measure of planarity $\calP$ in the $yz$-plane for ($a$) $U_0 = 0$ and ($b$) $U_0 = 1000$. Each plot is scaled individually between $\calP=1$ (white) and the minimum value of $\calP$ in the flow (black).}
\label{fig:non-planarity}
\end{figure}
\begin{figure}
\vskip -0.5 cm
\centerline{
		\includegraphics[scale = 0.6]{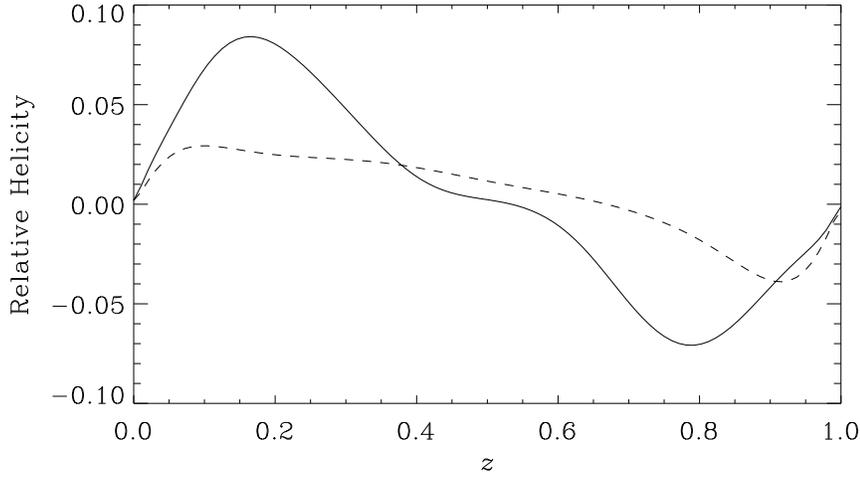}
	}
\caption{Snapshots of $h(z)$ (the horizontally averaged relative flow helicity) for $y < \lambda/2$ (dashed line) and $y > \lambda/2$ (solid line) for $U_0 = 1000$. Exact antisymmetry about the midplane ($z=0.5$) is recovered by time averaging.}
\label{fig:rel_hel}
\vskip -0.2 cm
\end{figure}

On increasing the amplitude $U_0$ of the target shear flow, various regimes can be identified in the nature of the resulting convection. These are demonstrated in figure~\ref{fig:ke}, which plots the ratio of the kinetic energy in the target flow to the total kinetic energy as a function of time for three different values of $U_0$, and figure~\ref{fig:temps}, which shows the corresponding density plots of the temperature fluctuations close to the upper boundary, together with that of the non-sheared state. For $U_0=300$ (shear parameter $S \approx 0.5$) the convection is such as to decrease the energy in the shear mode from its target value; note from figure~\ref{fig:ke}($a$) that, at least for this value of $U_0$, a long temporal integration is needed in order to determine the final stationary state. In figure~\ref{fig:temps}($b$) it is possible to detect a large-scale vortex underlying the small-scale convection. For $O(1)$ values of $S$, the kinetic energy in the shear flow is comparable with its target value, and this mode dominates the total kinetic energy (e.g.\ figure~\ref{fig:ke}($b$)) (It should be noted that the hydrodynamic state has been evolved for much longer than shown in figure~\ref{fig:ke}($b$), with no transition to a different state.)  As shown in figure~\ref{fig:temps}($c$), the shear leads to a clear elongation of the convective cells, together with significant inhomogeneity between the two halves of the domain in the $y$ direction. For $0<y<\lambda/2$, the vorticity augments the underlying vorticity due to the rotation of the layer, whereas for $\lambda/2<y<\lambda$ it tends to reduce it. The net underlying vorticity in the $z$-direction can be expressed in dimensionless form as
\begin{equation}
\Tay^{1/2}+\frac{2 \pi U_0}{\lambda} \sin \frac{2 \pi y}{\lambda} .
\end{equation}
Clearly (when $U_0$ is positive) the underlying vorticity has the smallest absolute value when $y=3 \lambda/4$. The vorticity dynamics in the two halves of the layer is similar if $U_0$ is very small or large; the maximum disparity between the two halves of the layer (in $y$) occurs when $U_0 \sim \lambda \Tay^{1/2}/2 \pi \approx 1125$ here. In figure~\ref{fig:temps}($c$), $U_0$ is close to this optimal value, and it can be seen that convection is indeed most vigorous in the neighbourhood of  $y=3 \lambda/4$. For $U_0 \lesssim 500$ ($S \lesssim 1$), convection dominates in the sense that there are no streamlines extending across the domain. For larger values of $U_0$ (e.g.\ figure~\ref{fig:temps}($c$)) a clear `channel flow' is established in $0<y<\lambda/2$. For a range of values of the shear amplitude $U_0$, this shear-dominated flow remains stable. However, at yet larger values of $U_0$, the shear becomes unstable and the resulting flow reverts to being less shear-dominated, as can be seen by figure~\ref{fig:ke}($c$) for $U_0=2000$. At these larger values of $U_0$, a large coherent vortex forms and the flow has a very different structure, with the convective cells expelled from the vortex (see figure~\ref{fig:temps}($d$)). In this paper we shall concentrate principally on the nature of the dynamo action resulting from values of $U_0$ for which the convection and shear flow can co-exist (e.g.\ figure~\ref{fig:temps}($c$)) and for which there is a clear separation in their spatial scales. Figure~\ref{fig:non-planarity} gives a measure of the planarity $\calP$ for the flows with $U_0 = 0$ and $U_0 = 1000$, where $\calP$ is defined as the ratio of the horizontal to total kinetic energies, 
\begin{equation}
\calP(y,z) = \frac{\langle \bfU_H^2 \rangle}{\langle \bfU^2 \rangle},
\end{equation}
with angle brackets denoting an average over $x$. It can be seen that for $U_0 = 1000$ the flow is essentially two-dimensional for much of the domain, with patches of fully three-dimensional flows centred around the turning points in the target shear flow. For Boussinesq convection, the helicity distribution is anti-symmetric about the mid-plane \citep[see, for example,][]{CS_72, CH_06}. However, the introduction of a shear flow in a rotating frame allows for differences between the domains $y < \lambda /2$ and $y > \lambda /2$; this is illustrated by figure~\ref{fig:rel_hel}, which shows the relative helicity $h(z)$ for the two halves of the $y$-domain, where
\begin{equation}
h(z)=\frac{\langle \bfu \cdot \nabla \times \bfu \rangle}{\langle \bfu^2 \rangle^{1/2}\langle (\nabla \times \bfu)^2 \rangle^{1/2}}\;,
\label{relhel}
\end{equation}
with the averages taken over horizontal planes. The helicity is significantly greater for $y > \lambda/2$, where the background and flow vorticities are of the same sign.
%------------
%------------
%------------
% Contours in yz (averaged over x) for KE, helicity, relative helicity, measures of non-planarity.

% Some contour plots to show differences in the flow in the two halves of the $y$-domain. 

\subsection{Kinematic Dynamo Action}

%------------
\begin{figure}
\vskip -0.5 cm
\centerline{
		\includegraphics[scale = 0.6]{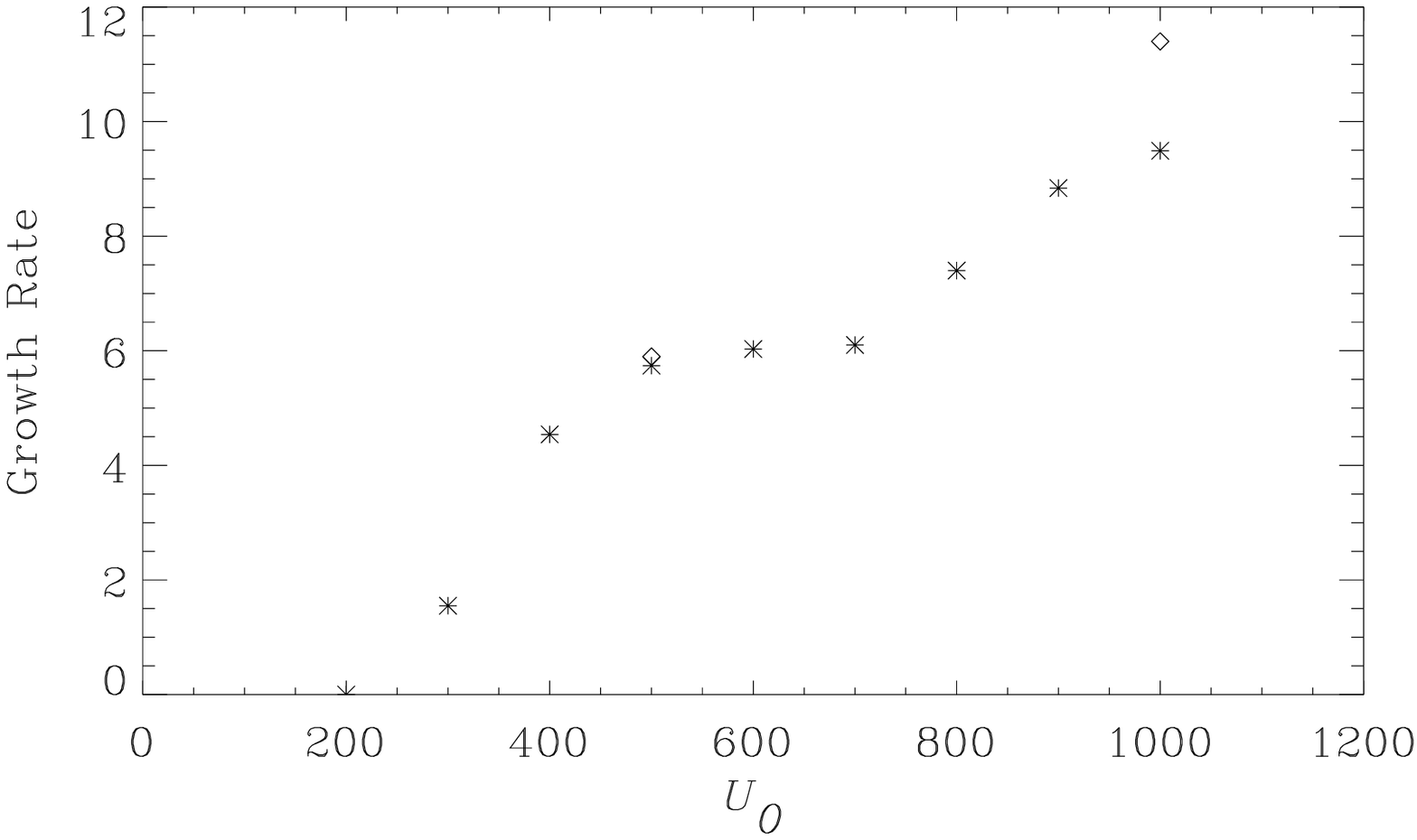}
	}
\caption{Dynamo growth rates versus $U_0$ for $\Ray=150\,000$ and $\lambda = 10$ (asterisks). The diamonds show the growth rates when the shear amplitude is $2U_0$  with $\lambda = 20$.}
\label{fig:growth_rates}
\vskip -0.5 cm
\end{figure}
%------------

Figure~\ref{fig:growth_rates} plots the dynamo growth rate as a function of $U_0$, for $\Ray=150\,000$. It can be seen that the incorporation of velocity shear facilitates dynamo action, with the critical value of the shear amplitude given by $U_0 \approx 200$ (i.e.\ $S \approx 0.3$). Further increases in $U_0$ serve to enhance the growth rate, although there is no simple power law relationship. The levelling off in the growth rate for $500 \lesssim U_0 \lesssim 700$ corresponds to a change in the nature of the flow regime, as described above. We have also calculated the dynamo growth rates for a domain that is twice as wide ($\lambda = 20$); a comparison between the domains of differing sizes then requires replacing $U_0$ with $2 U_0$, to keep the same value of $S$, according to definition~(\ref{eq:Sdef}). For the smaller value of $S$ there is little influence of the box size, whereas at the larger value of $S$, although the growth rates are similar, there is a clear influence of the domain size, with the dynamo in the larger domain being more efficient. In any case, we do not necessarily expect close agreement, since although the shear $S$ is the same for the $\lambda=10$, $U_0 =1000$ and $\lambda=20$, $U_0=2000$ runs, the convective structures are independent of the box size whereas the region of, say, positive shear scales with $\lambda$. The dynamo growth rates are also consistent with those found in \cite{HP_09} in a domain of half the width (so the values of $S$ in that paper should here be multiplied by two for comparison); the more extensive data that we now have makes it clear that the relationship between growth rate and shear is more complicated than the linear one with which the earlier data were consistent.

\begin{figure}
\vskip -2.5cm
\centerline{
		\includegraphics[scale = 0.6]{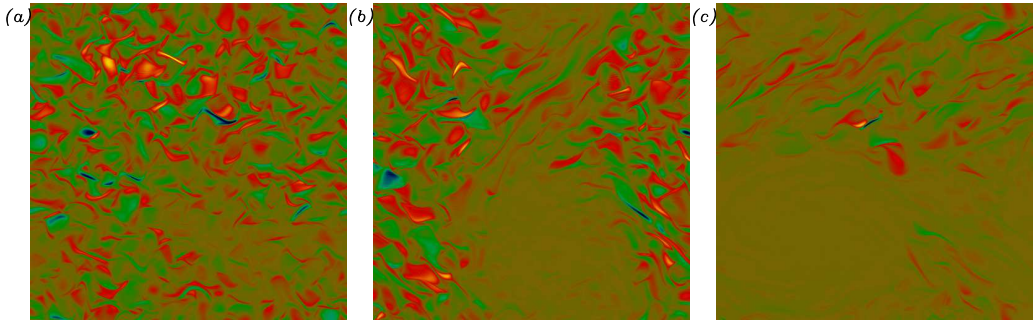}
	}
\caption{Density plots of $B_x$ at the upper boundary, for increasing values of $U_0$: ($a$) $U_0=200$, ($b$) $U_0=300$, ($c$) $U_0=400$. Each plot is scaled individually; colour table as in figure~\ref{fig:bxby} below.}
\label{fig:bx_diff_U0}
\end{figure}

Figure~\ref{fig:bx_diff_U0}, which plots $B_x$ at the top of the layer, illustrates how the magnetic field changes with increasing $U_0$. At $U_0 = 200$, essentially the smallest value of the shear flow that allows for dynamo action, although there is already some evidence of asymmetry between the two halves of the $y$-domain, there is still significant magnetic energy in $0 < y < \lambda/2$ (a $28\%$ to $72\%$ split in the energy of the $B_x$ field between the two halves of the $y$ domain). At $U_0 = 300$, the underlying vortex depicted in figure~\ref{fig:temps}($b$) has a clear influence, introducing a strong large-scale variation in magnetic field in the $x$-direction. With a further increase in $U_0$ to $U_0 = 400$, the flow enters the regime of $O(1)$ values of $S$, and the field shows a clear asymmetry between the two halves of the $y$-domain. The field persists in this form for a range of $U_0$, until the shear amplitude is sufficiently great as to trigger an instability (figure~\ref{fig:temps}($d$)).

%------------
%------------
\begin{figure}
\vskip -1.5 cm
\centerline{
		\includegraphics[scale = 0.6]{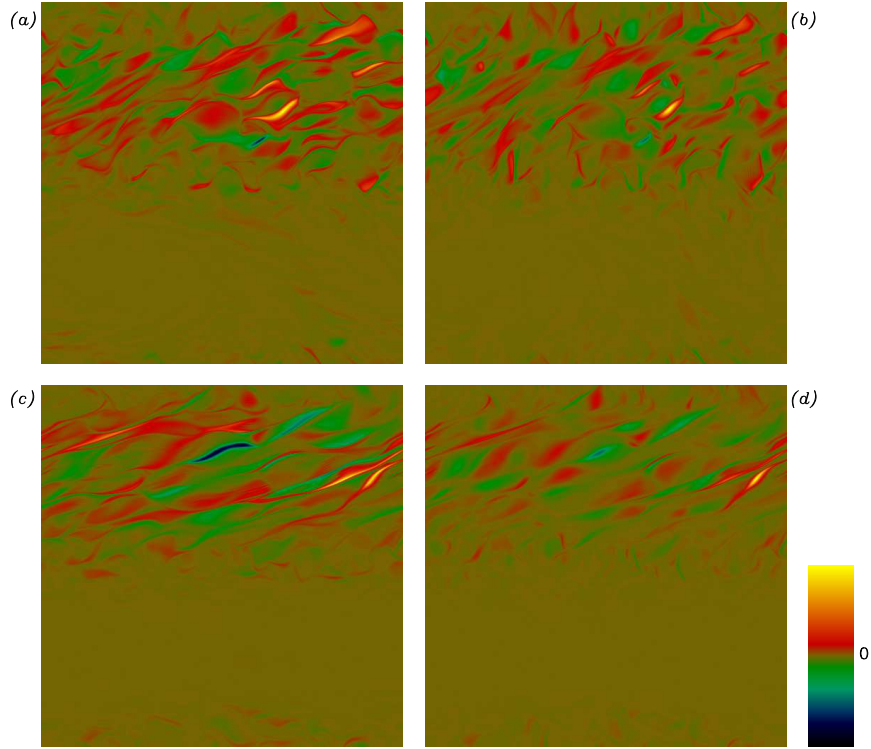}
	}
\caption{Snapshots of the two components of the horizontal magnetic field at the upper boundary: ($a$)~$B_x$ for $U_0=500$, ($b$)\ $B_y$ for $U_0=500$, ($c$)\ $B_x$ for $U_0=1000$, ($d$)\ $B_y$ for $U_0=1000$. The plots for each $U_0$ are scaled individually; the colour bar ranges from $-\textrm{max}|B_x|$ to $\textrm{max}|B_x|$.}
\label{fig:bxby}
\vskip -0.5 cm
\end{figure}
%------------

Figure~\ref{fig:bxby} shows snapshots of $B_x$ and $B_y$ in the $xy$-plane at the upper boundary, for $U_0 = 500$ and $U_0 = 1000$. A movie of such plots (included as supplementary material) reveals clearly the advection of the magnetic field pattern by the velocity shear. Two important features can be noted. One is that the dynamo action is strongly inhomogeneous, being concentrated in $\lambda/2 < y < \lambda$. The other is that the stronger shear leads to pronounced stretching of the field structures in the $x$-direction.

%------------
\begin{figure}
\vskip 0.2cm
\hskip 0.7cm
\centerline{
		\includegraphics[scale = 0.6]{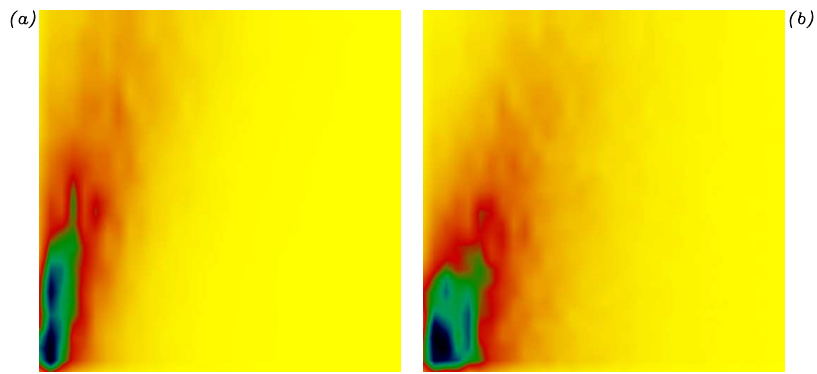}
	}
\caption{Density plots of the time-averaged and depth-averaged Fourier transform of the magnetic energy in ($a$)~$B_x$ and ($b$)~$B_y$ for $U_0 =1000$. The horizontal and vertical axes are the $x$ and $y$ wavenumbers, respectively, in the range $0 \le k < 32$. The plots are scaled individually, from blue (highest energy) to yellow (lowest).}
\label{fig:fft_me}
\end{figure}
%------------
\begin{figure}
\vskip 0.2cm
\centerline{
		\includegraphics[scale = 0.5]{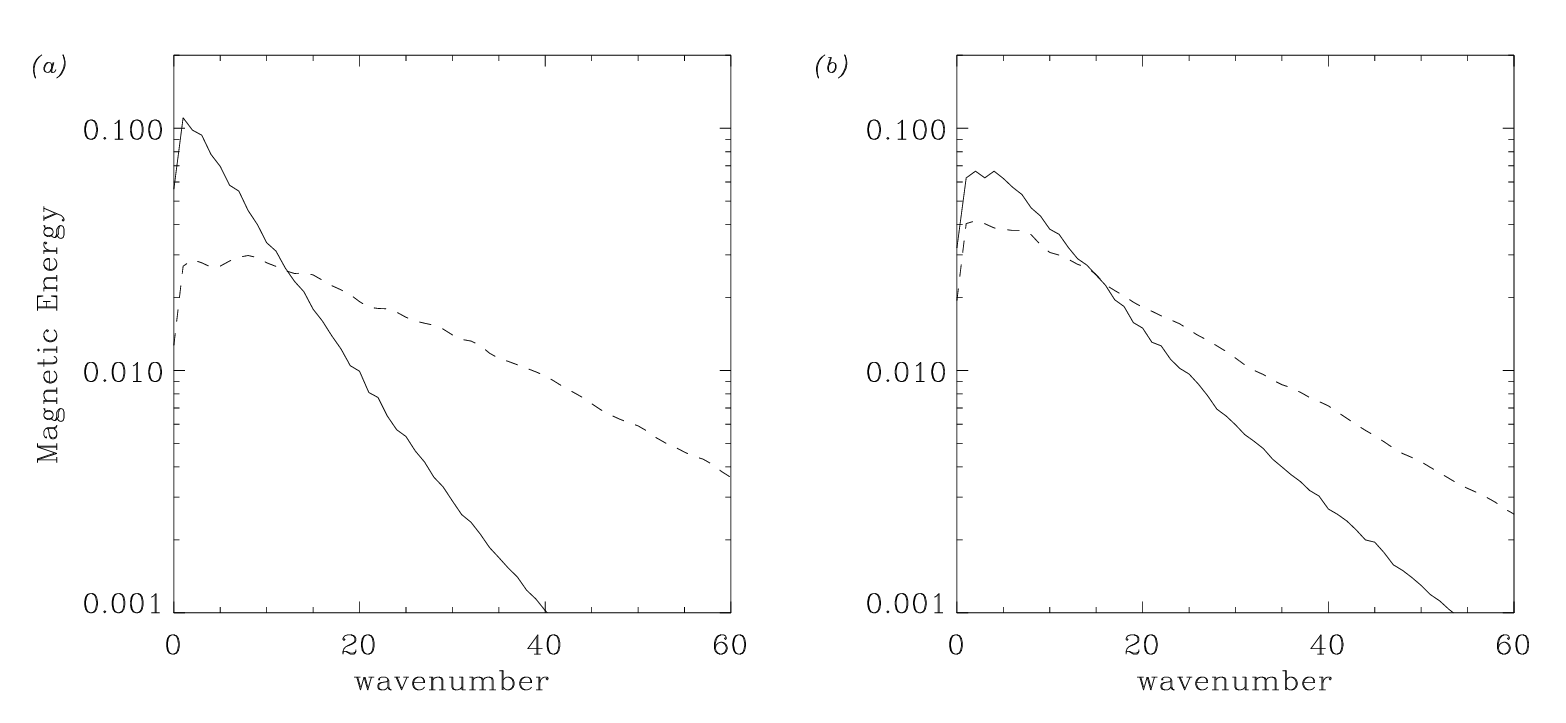}
	}
\caption{One-dimensional spectra of the energy in ($a$)\ $B_x$ and ($b$)\ $B_y$. The solid (dashed) lines show the spectra as a function of $k_x$ ($k_y$) after averaging over $y$ ($x$).}
\label{fig:mag-spectra}
\end{figure}
%------------

In order to obtain a quantitative description of the scales on which the field is being generated, two-dimensional Fourier transforms of $B_x^2$ and $B_y^2$ are constructed, after both depth averaging and time averaging and having removed the exponential growth of the field. Figure~\ref{fig:fft_me} shows these plots for the range of horizontal wavenumbers $0 \le k_x, k_y < 32$. The plots are scaled individually, but it should be pointed out that the bulk of the magnetic energy resides in $B_x$, as is maybe to be expected from a flow that is strongly influenced by a shear $U(y) \bfxhat$; for the parameter values of figure~\ref{fig:fft_me}, $\langle B_x^2 \rangle / \langle B^2 \rangle \approx 0.77$, $\langle B_y^2 \rangle / \langle B^2 \rangle \approx 0.14$, $\langle B_z^2 \rangle / \langle B^2 \rangle \approx 0.09$ (where angle brackets denote an average over the fluid volume and time). The distribution over wavenumbers displays a marked asymmetry in $k_x$ and $k_y$ for $\langle B_x^2 \rangle$, but is roughly symmetric  for $\langle B_y^2 \rangle$. For the former case, which provides the principal contribution to the overall magnetic energy, the dominant modes are $k_x = 1$, $k_y=1, 2, 3$; the fall off with energy with increasing $k_x$ is significantly greater than that with increasing $k_y$. An alternative representation of the distribution of the magnetic energy over wavenumbers is provided by figure~\ref{fig:mag-spectra}, which shows the one-dimensional spectra of $B_x^2$ and $B_y^2$ having summed over either $k_x$ or $k_y$. These spectra should be compared with that for the case of no velocity shear \citep[e.g.\ figure~6 of][]{CH_06}, in which the magnetic energy is peaked at the scale of the convective cells and falls off rapidly to both larger and smaller scales.

%Time traces of selected large-scale components of $B_x$, say --- from slices.

Having demonstrated the broad features of the dynamo-generated field, and shown that a field with large-scale structure in the horizontal plane is indeed produced, it is important to seek an understanding of the underlying physical processes responsible for field generation. In particular, is it possible to distinguish between the various scenarios outlined in \S\,\ref{sec:influence_shear}? To this end, in the following section we try to answer this question by comparing the dynamo properties of the actual convective flows with those of related flows obtained by the removal of selected Fourier modes, a process we term `filtration'.

\section{Filtered Flows} \label{sec:ff}

\subsection{The Filtration Process}

The process of filtration that we employ is essentially that of low- and high-pass filtration, first introduced into the study of turbulence by \cite{Obukhov_41}; in the context of isolating the important modes for dynamo action, the idea of spectral filtering has been explored by \cite{TC_08}.

By virtue of the periodicity in the $x$ and $y$ directions, all variables can be expressed as a sum of Fourier modes of the form 
\begin{equation}
f(z, k_x, k_y, t) \exp{\pm i (2\pi\lambda^{-1}(k_x x+k_y y))},
\end{equation}
where $k_x$ and $k_y$ are integers. If we denote a cut-off wave number by $k_{\rm cut}$ then the filtration takes one of the following forms:\\
(a)~short wavelength (SW) cutoff (i.e.\ long wavelengths retained): set to zero the amplitudes of all modes for which $k=\max(|k_x|,|k_y| ) >k_{\rm cut}$;\\
(b)~long wavelength (LW) cutoff (i.e.\ short wavelengths retained, plus the shear):  set to zero the amplitudes of all modes for which $k=\min(|k_x|,|k_y| ) <k_{\rm cut}$, {\it but retain the mode $(0,1)$ corresponding to the shear}. It is worth stressing that the amplitude of the $(0,1)$ mode emerges from the interaction of the convection with the imposed shear of amplitude $U_0$; for the flows considered in this section, the energy in the $(0,1)$ mode is about $90\%$ of the target energy.

The filtration is applied in $k_x$ and $k_y$ since we are addressing the issues of scale separation and large-scale field generation in the horizontal plane. For completeness, we have also performed some runs in which filtration has been applied to the vertical spectrum, but we have found that this further filtration makes no significant difference to the properties of the magnetic fields that result.

Whichever the filtration adopted, the procedure is as follows:\\
(1)~Solve the momentum and heat equations at full resolution;\\
(2)~At each time step perform the filtering to produce a filtered velocity $\bfu_{\rm f}$ together with the shear;\\
(3)~Solve the induction equation (\ref{eq:ind}) at full resolution with $\bfu$ replaced by $\bfu_{\rm f}$.

It is helpful to introduce some notation: thus SWC$n$ denotes a short-wave cutoff at $\kcut=n$ and LWC$n$ a long-wave cutoff at $\kcut=n$. In the following subsections we investigate the short- and long-wave cutoffs for a range of values of $\kcut$ for the case of $\Ray =150\,000$, $U_0=1000$.

\subsection{Short Wavelength Cutoffs}

%------------
\begin{figure}
\vskip -2cm
\centerline{
		\includegraphics[scale = 0.65]{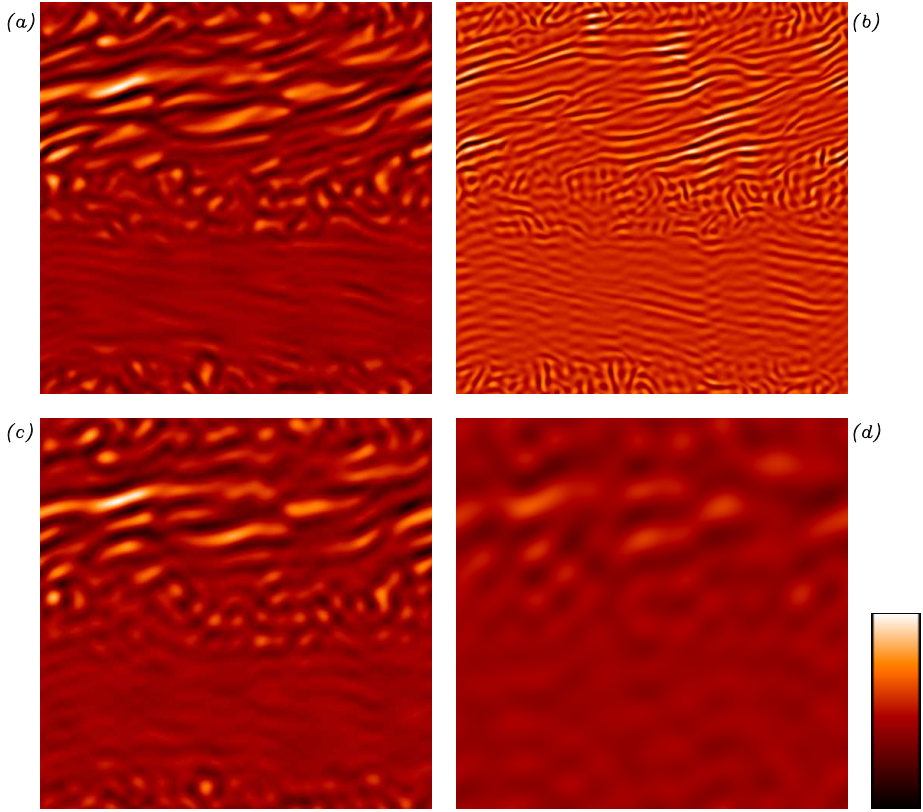}
	}
\caption{Density plots of the vertical velocity close to the upper boundary for ($a$)~SWC30, ($c$) SWC20 and ($d$) SWC10. The residual small-scale flow removed by the filtration SWC30 is shown in ($b$). Plots ($a$), ($b$), and ($c$) are on the same scale; the residual velocity in ($b$) is of much smaller amplitude and is scaled independently. White denotes upward velocity, black falling velocity.}
\label{fig:filtered_w}
\end{figure}
%------------
The influence on the convection of filtration via a short wavelength cutoff is exhibited clearly in plots of the vertical velocity, as shown in figure~\ref{fig:filtered_w}. As is to be expected, notable changes come about when $\kcut^{-1}$ is comparable with the number of convective cells across the domain. For the example shown in figure~\ref{fig:filtered_w}, there is little readily appreciable difference between the SWC30 flow and the full flow; the high frequency residual 
velocity that is removed by the filtration is shown in figure~\ref{fig:filtered_w}($b$).

Before discussing the influence of short wave cutoffs on flows with shear, it is important to understand the dynamo properties of an unsheared flow subject to the same filtration process. For the case of $\Ray = 150\,000$, which, recall, does not act as a dynamo in the absence of shear, the dynamo properties initially improve as $\kcut$ is decreased, owing to the removal of the damping effect of the small scales; indeed there is a range of $\kcut$ around $20$ for which the filtered flow acts as a dynamo. However, when $\kcut$ is sufficiently small so as to exclude the energy-containing modes of the convection, the filtered flow becomes too feeble to support dynamo action.

%------------
\begin{figure}
\vskip 0cm
\centerline{
		\includegraphics[scale = 0.65]{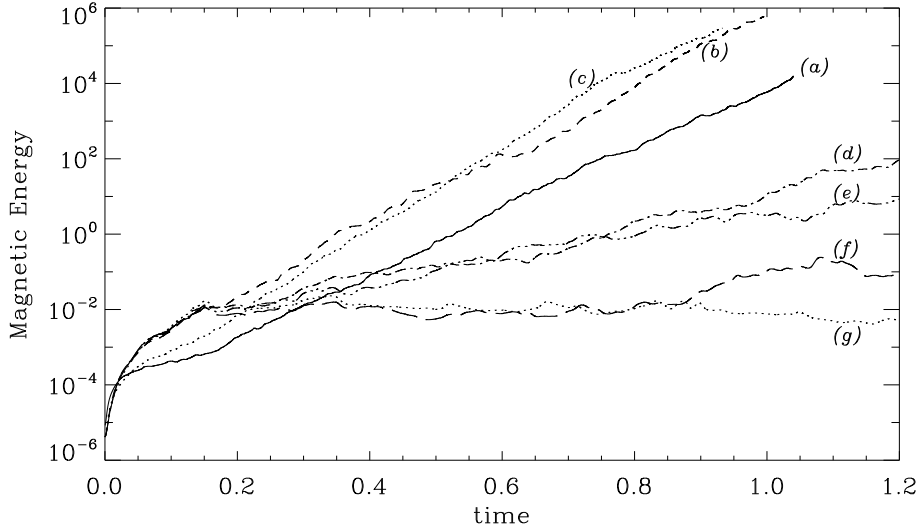}
	}
\caption{Magnetic energy versus time, with $U_0 =1000$ and for various shortwave cutoffs SWC$n$: ($a$)~all modes retained, ($b$) $n=20$, ($c$) $n=10$, ($d$) $n=5$, ($e$) $n=4$, ($f$) $n=3$, ($g$) $n=2$.}
\label{fig:me_vs_t_ssn}
\end{figure}
%------------

Analogous behaviour occurs in the sheared case, as shown in figure~\ref{fig:me_vs_t_ssn}, which plots the magnetic energy versus time for various filtrations with $U_0=1000$. As in the unsheared case, dynamo action is enhanced slightly by the removal of just the smallest scales. In comparison with the unsheared case, the value of $\kcut$ at which a significant reduction in the growth rate occurs is now rather smaller, reflecting the importance of modes of scale intermediate between that of the shear and that of the original (unsheared) convection; figure~\ref{fig:temps} clearly shows these longer scales. The SWC$n$ ($n=3,4,5$) flows are still able to support dynamo action, although the growth is weak and somewhat irregular. It is important to note that the dynamo fails if $\kcut =2$, thereby showing that the dynamo process does not depend on the largest velocity scales alone, but must rely crucially on velocity scales comparable with those of the convection.

%------------
\begin{figure}
\vskip -2.5cm
\centerline{
		\includegraphics[scale = 0.6]{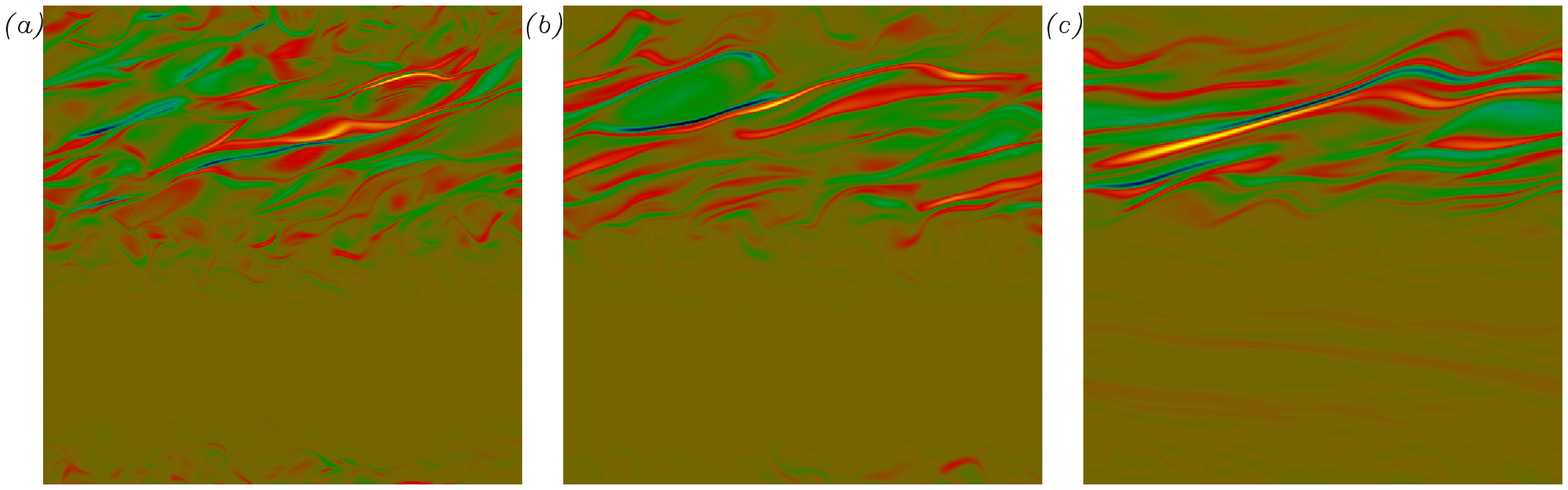}
	}
\caption{Density plots of $B_x$ at the upper boundary, for various shortwave cutoffs SWC$n$: ($a$) $n=20$, ($b$) $n=10$, ($c$) $n=5$. Colour table as in figure~\ref{fig:bxby}.}
\label{fig:bx_swc}
\end{figure}
%------------

The fact that the growth rate of, for example, the SWC20 and SWC10 dynamos is similar to that resulting from the full, unfiltered flow does not, of itself, mean that the same dynamo mode is being excited. It is, in addition, necessary to examine the morphology of the magnetic field generated. Comparison of figure~\ref{fig:bx_swc} with figure~\ref{fig:bxby}, which show the same localization and striated structures, does though confirm that the dynamo mechanism is identical in the filtered and unfiltered cases.

\subsection{Long Wavelength Cutoffs}

%------------
\begin{figure}
\vskip 0cm
\centerline{
		\includegraphics[scale = 0.65]{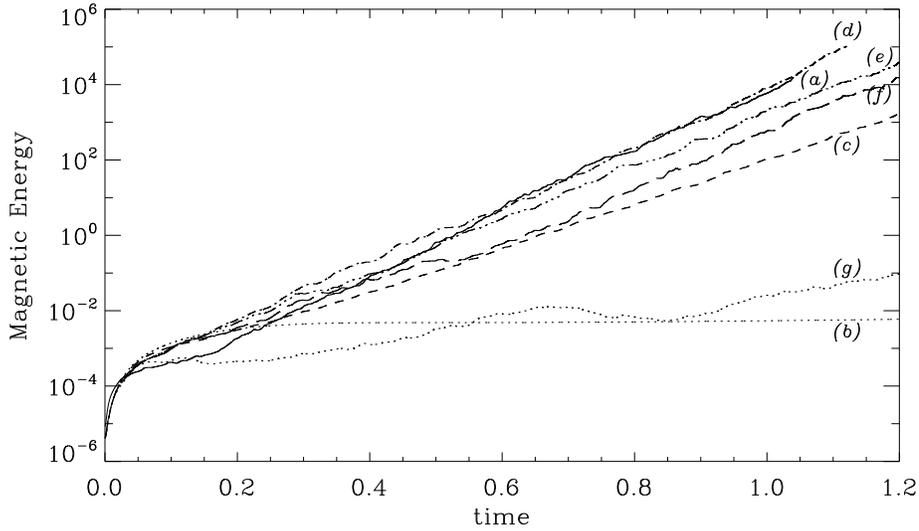}
	}
\caption{Magnetic energy versus time for various long-wave cutoffs LWC$n$: ($a$) all modes retained, ($b$) $n=20$, ($c$) $n=10$, ($d$) $n=5$, ($e$) $n=4$, ($f$) $n=3$, ($g$) $n=2$.}
\label{fig:me_vs_t_lsn}
\end{figure}
%------------

The idea behind implementing long wavelength cutoffs is to obtain velocity fields with a strict scale separation between the large scale of the shear and a much smaller convection scale; in this way we can find to what extent the dynamo is of a classical mean field $\alpha \omega$ type. It should first be noted that there is no dynamo action resulting from flows with long-wave cutoffs that also discard the target shear mode. Although perhaps not too surprising, this rules out the possibility that the influence of the shear on the small scales is such that the small scales, of themselves, become capable of acting as a dynamo. 

Figure~\ref{fig:me_vs_t_lsn} shows the temporal growth of the magnetic energy for LWC$n$ flows for a range of values of $n$. For sufficiently large $n$, the small scales retained, in conjunction with the shear mode, do not support dynamo action. However, the $n=20$ flow does act as a dynamo, with a very well defined uniform growth of the field with time, albeit with a slow growth rate. As can be seen from figure~\ref{fig:me_vs_t_lsn}, decreasing $n$ further leads to more efficient dynamo action, with the dynamo growth rate of the LWC$5$ flow becoming comparable with that of the full convective flow. It is surprising that dynamo action for the LWC$2$ flow is much weaker than that for the full flow; it is though possible to shed some light on this by consideration of the structure of the magnetic field generated.

%------------
\begin{figure}
\vskip -2.5cm
\centerline{
		\includegraphics[scale = 0.6]{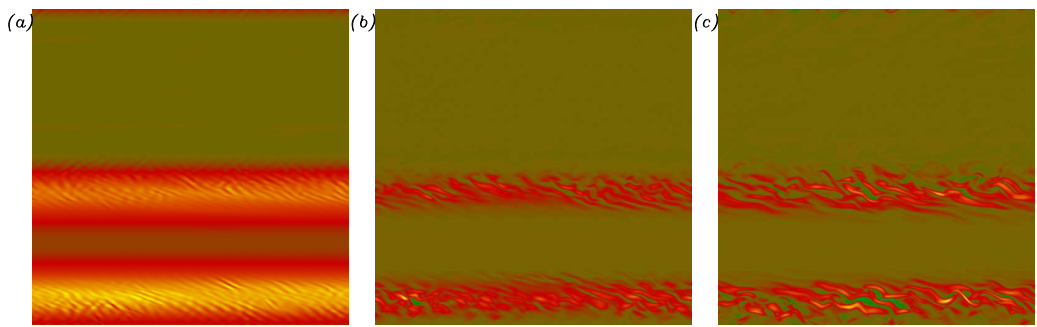}
	}
\caption{Density plots of $B_x$ at the upper boundary, for various long-wave cutoffs LWC$n$: ($a$) $n=20$, ($b$) $n=10$, ($c$) $n=5$. Colour table as in figure~\ref{fig:bxby}.}
\label{fig:bx_lwc}
\end{figure}
%------------

Simply from inspection of the growth rate, we cannot rule out the possibility that the dynamo mechanisms of the full flow and the LWC flows are the same; for example, the growth rates of LWC$10$ and LWC$5$ are very similar to that of the full flow. However in this case, examination of the magnetic field reveals that of the LWC dynamos to be very different in spatial structure from that resulting from the full flow. Figure~\ref{fig:bx_lwc} shows $B_x$ at the upper boundary for three LWC flows. In marked contrast to the magnetic field generated by the full flow (figure~\ref{fig:bxby}), the field is essentially zero in $\lambda/2<y<\lambda$ and is confined to two bands in $0<y<\lambda/2$.  It is of interest to note that magnetic field is expelled from regions of vigorous flow in $\lambda/2 < y < \lambda$ and from the region of strong shear in  $0<y<\lambda/2$. Figure~\ref{fig:bx_lwc} exhibits a large-scale modulation in $y$ in addition to small-scale fluctuations. Therefore it might be thought that such a dynamo could be understood within the mean field framework with an averaging that allows for $y$ modulation. However, direct calculation of the $\alpha$-effect reveals that $\alpha$ is large where the field is weak and so it is not clear that such a mean field description is appropriate. The anomalous behaviour of the LWC$2$ flow in figure~\ref{fig:me_vs_t_lsn} can be understood in terms of the existence of two different types of dynamo mechanism. The field generated by LWC$2$ is of the type exhibited in figure~\ref{fig:bx_lwc} for the higher LWC$n$ modes, and for which the addition of low $n$ modes is eventually detrimental to dynamo action. It is only when the $n=1$ modes are included that the true dynamo is recovered.

\section{Higher Rayleigh Number} \label{sec:hrn}

For $\Ray \gtrsim 170\,000$ (when $\Tay = 500\,000$), the convective flow supports kinematic dynamo action even in the absence of an imposed shear flow; as shown by \cite{CH_06}, the generated field is small-scale. The standard formulation of mean field dynamo theory proceeds on the assumption that small-scale dynamo action is not sustainable and that any small-scale field results only from the interaction between a large-scale field and a small-scale velocity. Although here the situation is somewhat different in terms of the description of the large-scale magnetic field, it is nonetheless of interest to explore whether the results of \S\S\ref{sec:ffrsc},\ref{sec:ff} are critically dependent on the lack of dynamo action in the absence of shear, or if they are more widely applicable. Here we concentrate on the case of $\Ray = 250\,000$.

%------------
\begin{figure}
\vskip -0.5 cm
\centerline{
		\includegraphics[scale = 0.6]{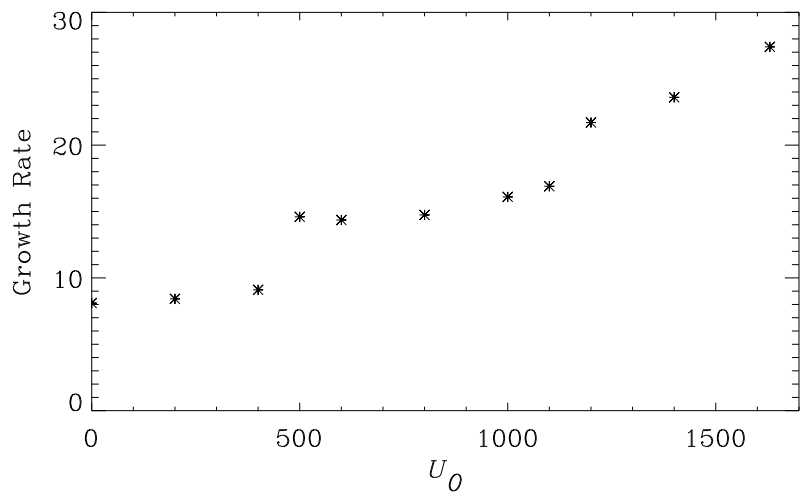}
	}
\caption{Dynamo growth rates versus $U_0$ for $\Ray=250\,000$ and $\lambda = 10$.}
\label{fig:growth_rates_ra_250}
\vskip -0.5 cm
\end{figure}
%------------

\begin{figure}
\vskip -1cm
\centerline{
		\includegraphics[scale = 0.7]{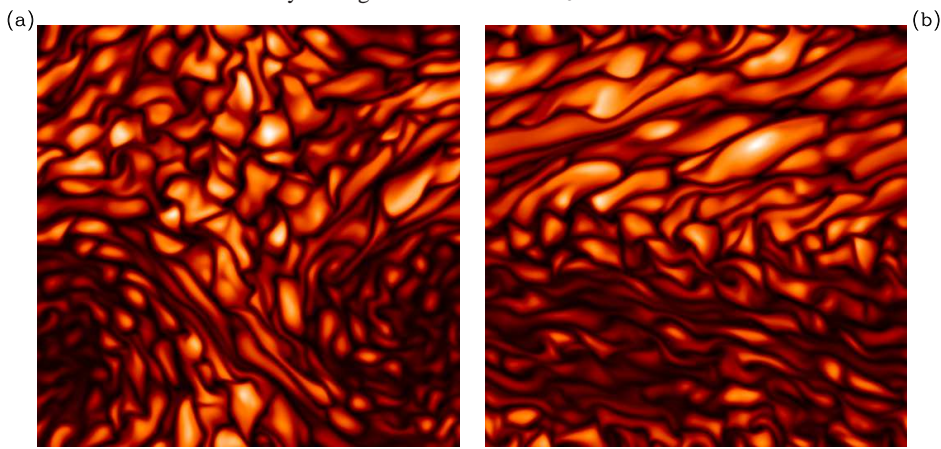}
	}
\caption{Snapshots of the temperature perturbations close to the upper boundary for $\Ray = 250\,000$ and two different values of the shear flow: ($a$) $U_0 = 600$, ($b$) $U_0 = 1630$.}
\label{fig:temps_ra_250}
\end{figure}

Figure~\ref{fig:growth_rates_ra_250} plots the dynamo growth rate versus $U_0$ for $\Ray=250\,000$. Here the higher Rayleigh number leads to a greater kinetic energy in the absence of the shear flow, leading to $S \approx U_0/1000$. The largest value of $U_0$ shown is $U_0=1630$, which has the same value of the shear parameter $S$ as the flow with $U_0=1000$ at $\Ray=150\,000$. The incorporation of shear is again destabilising, with a similar, non-straightforward, dependence of the growth rate on $U_0$ as exhibited at the lower Rayleigh number (cf.\ figure~\ref{fig:growth_rates}). Three different regimes can be identified. For $U_0 \lesssim 400$ ($S \lesssim 0.4$), the shear is not sufficiently strong to change the basic convection pattern and hence the resulting dynamo action. For $500 \lesssim U_0 \lesssim 1100$, the small-scale convection is modulated by large-scale vorticity, as shown in figure~\ref{fig:temps_ra_250}$(a)$. For $U_0 \gtrsim 1200$, the convective pattern clearly reflects the influence of the target shear flow, as shown in figure~\ref{fig:temps_ra_250}$(b)$. Note that although figure~\ref{fig:temps}$(c)$ and figure~\ref{fig:temps_ra_250}$(b)$ show flows with the same formal value of the shear parameter $S$, and are indeed similar in structure, the shear at the lower $\Ray$ is slightly more dominant in stretching out the convective cells in $0 < y < \lambda/2$.

%------------
%------------

\begin{figure}
\vskip -2.5cm
\centerline{
		\includegraphics[scale = 0.6]{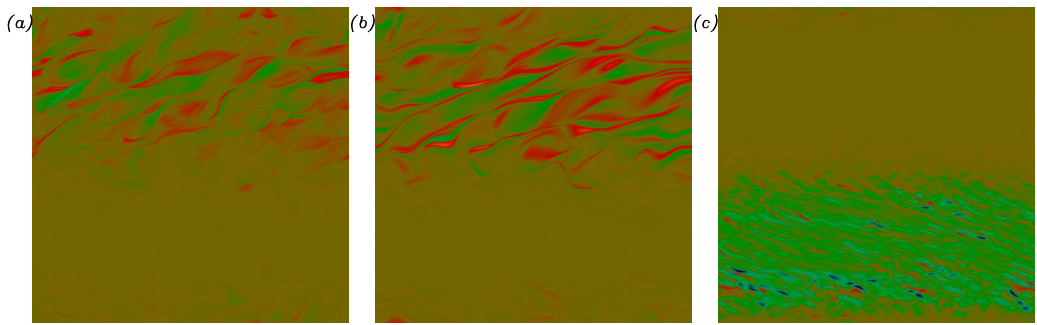}
	}
\caption{Density plots of $B_x$ at the upper boundary, for $\Ray =250\,000$: ($a$) all modes retained, ($b$)~SWC$10$, ($c$)~LWC$10$. Colour table as in figure~\ref{fig:bxby}.}
\label{fig:bx_ra_250}
\end{figure}

Comparison of the growth rate dependencies and convective flow patterns for the two different Rayleigh numbers suggests that, once the shear flow is influential, the underlying dynamo mechanism is the same in the two cases; the fact that the flow acts as a dynamo at the higher $\Ray$ in the absence of shear would therefore appear immaterial. To confirm this it is though also necessary to look at the structure of the magnetic field generated at the higher value of $\Ray$ and to verify that the effects of the filtration process are similar to those discussed in \S\ref{sec:ff}. As for the lower Rayleigh number example, we have examined the nature of the dynamo action resulting from short and long wavelength cutoffs of the flow. The overall trend is found to be the same. For the short wavelength cutoffs SWC$n$, the dynamo growth rate first increases as $n$ decreases, is maximized at some $n$ in the range $10 < n < 20$, and then decreases rapidly for $n<5$; thus a range of spectral modes is required for efficient dynamo action. As shown in figures~\ref{fig:bx_ra_250}$(a, b)$, the magnetic field generated by the short wavelength cutoffs is consistent with that generated by the entire flow. For long wavelength cutoffs LWC$n$, dynamo action ensues for $n \lesssim 30$ (a somewhat higher value of $n$ than for the $\Ray = 150\,000$ case) and the growth rate initially increases as $n$ decreases; again though, as can be seen from figure~\ref{fig:bx_ra_250}$(c)$, the resulting magnetic field is of a very different form to that generated by the full flow.

%------------
%------------

\section{Discussion}\label{sec:disc}

The research reported in this paper has allowed us to gain a full understanding of the phenomenon first described in \cite{HP_09}, which indicated that the incorporation of a large-scale shear flow into rotating convection promoted the generation of large-scale magnetic fields. We have examined the importance for field generation of the various scales in the flow by considering spectrally filtered velocity fields. Our main result, which we believe to be potentially significant in terms of understanding astrophysical magnetic field generation, is that the observed dynamo process depends for its existence on the entire range of scales from the shear flow down to the scale of the convective cells. In \S\,\ref{sec:influence_shear} we speculated that the introduction of shear might enhance the efficacy of the two-scale (mean field) dynamo process for turbulent, high $Rm$ flows. Instead we see that the dynamo is produced by a completely different mechanism with no scale separation. It thus remains an open question as to whether the mean field \textit{ansatz} is ever appropriate in these circumstances.

Our first aim was to confirm the earlier result that the addition of shear to a convective flow led to an enhancement of dynamo action. The growth rate of the dynamo increases with the magnitude of the shear $S$. Numerical studies of forced turbulence with shear by \cite{yousef2008a, yousef2008b}, together with a calculation for a simple model by \cite{hein2011}, suggest that over a range of shear amplitudes the growth rate of such a dynamo should increase linearly with the shear, though \cite{Proc12} has shown analytically that the linear scaling cannot continue to arbitrarily large values of $S$. In the present case, for both values of the Rayleigh number studied, the growth with $S$ is far from linear. This is because the nature of the flow changes significantly as the shear is increased, as discussed in \S\,\ref{subsec:isc}. 

We considered two different values of the Rayleigh number. For $Ra=150\,000$, the convective flow is not a dynamo, and so the effect of shear is crucial. For $Ra=250\,000$, on the other hand, there is a small-scale dynamo even in the absence of shear. Adding shear to this flow enhances dynamo action and produces a large scale component to the magnetic field, but the morphology of the evolving fields is very similar when $S=O(1)$. This suggests the interpretation that the action of the shear, whether on a small-scale `non-dynamo' or a small-scale dynamo, is very similar in both cases.

While \cite{yousef2008b} considered forced rotating turbulence with shear, the earlier paper \citep{yousef2008a} has no superimposed rotation, and yet the dynamo appears to function in a very similar manner in both cases. This led us to undertake further computations to look at the effects of shear on dynamo action in a non-rotating layer. However, in the absence of rotation, the imposed shear turns out to be readily destabilized by the convective flow and, at least with our target shear flow, it was not possible to attain a stable state with $O(1)$ values of the shear parameter $S$. The same problem precluded any systematic investigation of the `shear-current effect' \citep[see, for example,][]{RogKle2007, SridSingh2010}.

We have tried to understand the nature of the dynamo process by considering `filtered' flows. Clearly the dynamo is not much influenced by the smallest scales of flow, but removing scales intermediate between the shear and the turbulence has a huge effect on the form of the growing field, but not necessarily on the growth rate. One might expect a true mean field dynamo in the latter case, but in fact the largest scale of variation of the field is much smaller than the scale of the shear and, although a reduced model might be constructed by averaging along the direction of the shear flow, the resulting emf cannot be represented by a mean-field coefficient of the usual kind. So, paradoxically, the creation of conditions for a mean field dynamo precludes a dynamo of mean field type! All scales of flow except the smallest are needed to describe the dynamo process that is observed. To date, we have only considered a simple filtering process in Fourier space, isotropic in the horizontal directions. Given that anisotropy is introduced by the shear flow, it would be interesting to consider filtrations for which $\kcut$ is different in the $x$ and $y$ directions. More broadly, further physical insights may be gained by employing a wavelet filtration, where one could filter in space as well as in scale of variation.

There are of course other ways of combining shear and convection to produce a dynamo; for example, shear can be created through an Ekman layer in a rotating convecting fluid \citep[e.g.][]{ponetal01, zgz06}; alternatively, the shear might be produced as a thermal wind by horizontal temperature gradients. It would be of interest to know what scales of motion control the appearance of the dynamo in these cases.

The present study has investigated only the kinematic phase of the dynamo; the effects of the Lorentz force on the flow, which will eventually lead to equilibration, have been ignored. The final form of the magnetic field and the relation of the shear amplitude to the final magnetic energy are, however, of considerable interest. Our earlier results \citep{HP_09} suggest that, somewhat surprisingly, the final field amplitude is almost unaffected by the shear provided that the dynamo is sufficiently vigorous. We intend to return to this question in future work.

\begin{acknowledgments} 

We are grateful to Profs. C.A.\ Jones and S.M.\ Tobias for useful discussions. We should also like to thank the referees, whose helpful comments improved the presentation of the paper. The research was supported by STFC, and the computations were performed on the STFC-funded UKMHD parallel cluster.

\end{acknowledgments}

\end{document}